# Time Distributions of Large and Small Sunspot Groups Over Four Solar Cycles

A. Kilcik[1], V.B. Yurchyshyn[1], V. Abramenko[1], P.R. Goode[1],
A. Ozguc[2], J.P. Rozelot[3], W. Cao[1]

[1] Big Bear Solar Observatory, Big Bear City, CA 92314 USA
[2] Kandilli Observatory and Earthquake Research Institute, Bogazici University, 34684 Istanbul Turkey
[3] Nice University, OCA-Fizeau Dpt. Av. Copernic, 06130 Grasse France

Abstract: Here we analyze solar activity by focusing on time variations of the number of sunspot groups (SGs) as a function of their modified Zurich class. We analyzed data for solar cycles 20-23 by using Rome (cycles 20-21) and Learmonth Solar Observatory (cycles 22-23) SG numbers. All SGs recorded during these time intervals were separated into two groups. The first group includes small SGs (A, B, C, H, and J classes by Zurich classification) and the second group consists of large SGs (D, E, F, and G classes). We then calculated small and large SG numbers from their daily mean numbers as observed on the solar disk during a given month. We report that the time variations of small and large SG numbers are asymmetric except for the solar cycle 22. In general large SG numbers appear to reach their maximum in the middle of the solar cycle (phase 0.45-0.5), while the international sunspot numbers and the small SG numbers generally peak much earlier (solar cycle phase 0.29-0.35). Moreover, the 10.7 cm solar radio flux, the facular area, and the maximum CME speed show better agreement with the large SG numbers than they do with the small SG numbers. Our results suggest that the large SG numbers are more likely to shed light on solar activity and its geophysical implications. Our findings may also influence our understanding of long term variations of the total solar irradiance, which is thought to be an important factor in the Sun – Earth climate relationship.

1. Introduction

The Sun is a complex and dynamic body, releasing energy on all wavelength intervals of the electromagnetic spectrum. The changes in the output energy can be traced easily from many solar activity indicators, such as the international sunspot number (ISSN), sunspot areas (SSA), total solar irradiance (TSI), and 10.7 cm solar radio flux (F10.7). All of these indicators show cyclic behavior from days to hundreds of years (Krivova & Solanki 2002; Broun et al. 2005; Atac et al. 2006, Kilcik et al. 2010). Due to their particularly long temporal extension (~400 years), the ISSN and SSA are the most used solar activity indicators, many of them are strongly correlated with each other (Hathaway et al. 2002). The length of the solar cycle varies between 7.4 and 14.8 years and often depends on the method of determination (Usoskin & Mursula, 2003). The monthly sunspot numbers show a strong anti-correlation (–0.82) with the ascending time of the solar cycle or the length of the cycle (Waldmeier 1939, Bai 2006, Kilcik et al. 2009), the sunspot numbers are usually calculated as daily Wolf numbers:

$$R_z = k(10g+f), \qquad (1)$$

where f is the number of individual spots, g is the number of sunspot groups, and k is a correction factor, which varies with location and instrumentation (also known as the observatory factor or the personal reduction coefficient). As it follows from Eq. 1, the sunspot numbers are directly related to the number of SGs that are present, respectively on the solar disk on a given day.

The Zurich sunspot group classification was first introduced by Waldmeier (1947). The Zurich classification emerged as a modified classification scheme introduced by Cortie (1901). It was later modified by McIntosh (1981), and this modified Zurich classification is based on three components. The first component is the sunspot group class (used in the present paper), the second component describes the largest spot in a group, and the third one is the degree of spottedness in the group's interior (for more detail see, McIntosh 1990). Both original and modified Zurich classifications have the same first classification parameter, which is the group class. In the modified Zurich classification, the definitions of the classes were adopted from the Zurich classification, with the one exception that G and J classes were removed from the modified classification.

Various studies show a well–pronounced relationship between solar flare activity and coronal mass ejections (CMEs) on one hand, and complexity of sunspot groups (Waldmeier 1938, McInthosh 1990, Abramenko 2005, Bai 2006, Abramenko & Yurchyshyn 2010a, Abramenko & Yurchyshyn 2010b, and reference therein). On the other hand, solar cycle 23, however, shows some differences as compared to the previous two solar cycles. De Toma et al. (2004) found that the total solar irradiance (TSI) did not change significantly from cycle 22 to cycle 23 despite the decrease in sunspot activity of cycle 23 relative to cycle 22. Gopalswamy et al. (2003) argued that the CME occurrence rate in cycle 23 peaked two years after the solar cycle maximum, while Abramenko et al. (2010) reported that many low latitude coronal holes were observed during the declining and minimum phase of solar cycle 23. Kilcik et al. (2011) analyzed the maximum speeds of CMEs and found that they peaked approximately two years after the solar cycle maximum. These few examples clearly demonstrate that there are still many unanswered questions about the behavior of the 11-year solar cycle.

To answer some of these questions, here we focus on a comparison of two types of sunspot groups throughout solar cycles 20–23. In our analysis, sunspot groups of A, B, C, H, and J Zurich classes (the simplest groups) belong to the first type. The second type includes D, E, F, and G Zurich classes, which are the most complex active regions.

2. Data, Comparisons and Results

In this study, we investigate time distribution of SGs depending on their class. The SG classification data are only available for the last 4 solar cycles, thus the analyzed time interval only includes solar cycles 20 through 23. Here, the Zurich classification (cycle 20, 21) and the modified Zurich classification (cycle 22, 23) SG data were used. In general, both classification schemes are similar and the main difference between them is that the modified Zurich classification does not include classes G and J, present in the original scheme. Instead, sunspot groups of class G and J in the modified classification belong to classes D, E, F and H, correspondingly. Therefore chose of either of these two different schemes does not influence our results significantly. All data sets used in this study were taken from National Geophysical Data Center (NGDC).

The F10.7 cm radio flux measurements were obtained at the Dominion Radio Astrophysical Observatory, National Research Council of Canada. The used radio flux measurements represent the solar radio flux per unit frequency at a wavelength of 10.7 cm.

These SG classification data were collected by United States Air Force/Mount Wilson (USAF/MWL) observatory and the available online at the NGDC web page[1]. The USAF/MWL database includes measurements from Ramey Solar Observatory (RAMY; 18N, 67W), Holloman Solar Observatory (HOLL; 33N, 106W), Palehau Observatory (PALE; 21N, 158W), San Vito Solar Observatory (SVTO; 41N, 18 E), Learmonth Solar Observatory (LEAR; 22S, 114E), and others. For solar cycles 22 and 23, we used the LEAR station data as the principal data source, which had the best coverage for those cycles. All LEAR data gaps were filled with data from the other stations listed above, so that nearly a continuous time series of daily number of SGs was produced. For solar cycles 20 and 21, we used the only available, which was from Rome Astronomical Observatory (ROME, 42N, 12 E) SG data. In Figure 1, we show the coverage and data gaps in the final time series by plotting the total number of observing days per month during a solar cycle. As it follows from the figure, solar cycle 21 is the least well–covered and has more small data gaps than any other cycle. On the other hand, data sets for solar cycles 22 and 23 are the most reliable and they have nearly 100% coverage during the solar maximum period, which is the focus of this study. There were also large data gaps: solar cycle 20 data has a four months gap from January to April of 1972, solar cycle 21 data has a three months gap between January and July of 1984 and one month worth of missing data in June 1986. Finally, there are no available data for the entire year of 1995 (solar cycle 22). There are strong fluctuations in the numbers of observing days especially during the minimum phase of all investigated cycles. These missing daily records may be either due to weather conditions, technical problems, the absence of sunspots, or all of the above. We note that all large data gaps occurred either in the descending phase of the solar cycle or near the solar minimum. Therefore, they do not affect our results significantly.

---

[1] ftp://ftp.ngdc.noaa.gov/STP/SOLAR_DATA

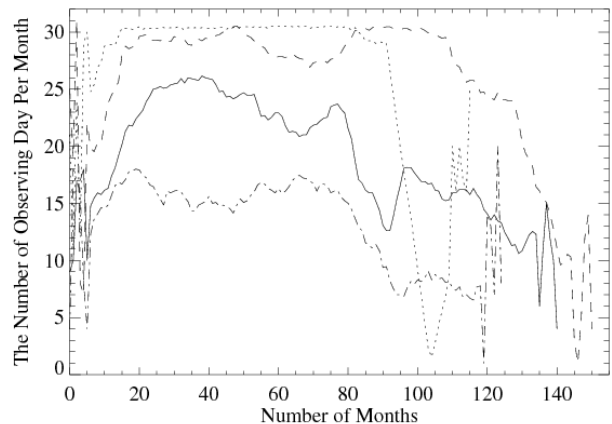

*Figure 1. Data coverage separately plotted for each solar cycle. Each curve shows the number of observing day per month. The data were smoothed with a 12 step running average filter. The solid/dot-dashed/dotted/dashed line is for cycle 20/21/22/23, respectively.*

For every SG in the data set, a classification according to the Zurich (solar cycle 20 and 21) or modified Zurich (solar cycle 22 and 23) classification, was assigned. It is possible to separate them into two distinct types. The first type consists of "large" SGs of classes D, E, F and G. In this group, the main leading and following spots should be present, and they should have a penumbra. Their longitudinal extent is generally larger than 5 degree. Thus, this type contains the majority of complex active regions. The second type includes "small" SGs that constitute A, B, C, H, and J classes. Here the requirement is that at least one main sunspot should have a penumbra (classes C, J, and H) or no penumbra (classes A and B; for more detail, see Waldmeier 1947; McIntosh 1990).

To mitigate the effect of data gaps, we derived the mean daily SG number by first calculating the total number of SGs of a given class observed during one month and then dividing this sum by the total number of observing days during this month. We thus obtained a parameter, essentially independent of data gaps. To remove short term fluctuations and reveal long term trends, the data were smoothed with a 12 month running average filter.

As we mentioned above, solar cycles 20 and 21 were substantially less well–covered as compared to cycles 22 and 23. Since here we are mainly interested in the general trend of the time distribution, rather than the absolute magnitude of the mean daily SG numbers, we first test how different data coverage effects the time distributions of SGs. In Figure 2, we plot the mean daily SG numbers as determined from ROME and LEAR data for the time interval between December 1981 and December 1985. For better comparison, ROME SG numbers were scaled to match the LEAR data. As it follows from the figure, the time variations of the large SG numbers, as derived from ROME and LEAR stations, are nearly identical (compare solid and dotted curves). The scaling coefficient for the ROME data is 1.2 and the cross-correlation coefficient between them is 0.99±0.006. The small SG numbers exhibit a larger difference in magnitude (scaling coefficient is 1.6, solid and dotted lines with symbols), while their time variations are similar, too. The cross-correlation coefficient in this case is also high, 0.99±0.009. We thus conclude that the data from these two stations can be used to analyze time variations of SG numbers during cycles 20-23, however, we shall keep in mind that the ROME SG numbers (cycles 20 and 21) are systematically underestimated as compared to LEAR data (cycles 22 and 23).

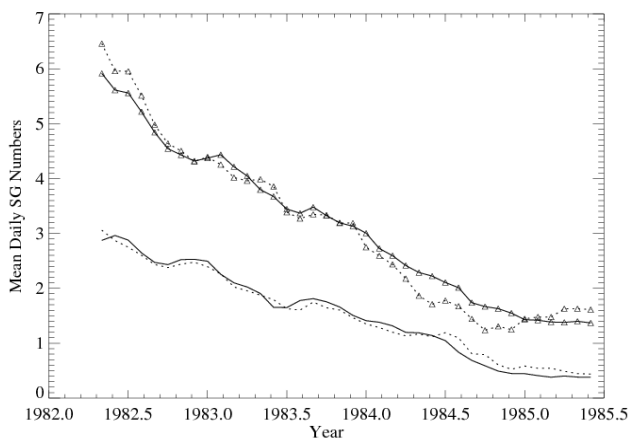

*Figure 2. Mean daily* SG numbers *as determined from LEAR (solid lines) and ROME (dotted lines) data. Lines with (without) symbols represent small (large) SG numbers. ROME small* SG numbers *were multiplied by 1.6 to match the LEAR data, while ROME large* SG numbers *were multiplied by 1.2 to match LEAR measurements.*

In Figure 3, we plot time variations of the SG numbers, the ISSN and the radio flux for four solar cycles. There are several interesting points that we would like to address. First, while large SG numbers show typical solar cycle variations in four profiles, they are also different. Second, it appears that the large SG numbers are lower during cycles 20 and 21 than those for cycles 22 and 23. As we mentioned earlier, cycle 20 and 21 data were taken from the ROME station, which has more data gaps compared to LEAR station (see Figure 1 and 2). Figure 2 also shows that the ROME and LEAR large SG numbers can be satisfactorily matched if the ROME data were scaled by 1.2. Taking into account these scaling coefficients, we may speculate that the large SG numbers in cycle 21 are the same or even higher compared to cycle 22 (we will return to this comparison in the Discussion section).

On the other hand, we note that the ISSN data display a double maximum for cycles, except cycle 20 has only one broad maximum (Figure 3). Also, the small and large SG numbers show similar time variations only during cycle 22 (years 1986-1996). Cycles 20, 21 and 23 show quite different behavior: while the small SG numbers peak during the first maximum of the ISSN, the large SG numbers peak at the second maximum of ISSN. Finally, the F10.7 radio flux tends to follow time variations of the large SG number. Note that in cycle 22 (which is the shortest cycle in the investigated time period) all these parameters behave in the same manner without displaying any noticeable difference between small and large SG numbers.

In Figure 4, we show the SG numbers overplotted with the measurements of facular area for cycles 22 and 23, (San Fernando Observatory, Chapman et al. 2004). The facular area appears to follow the large SG numbers rather than the small SG numbers (see Table 1 for cross-correlation coefficients).

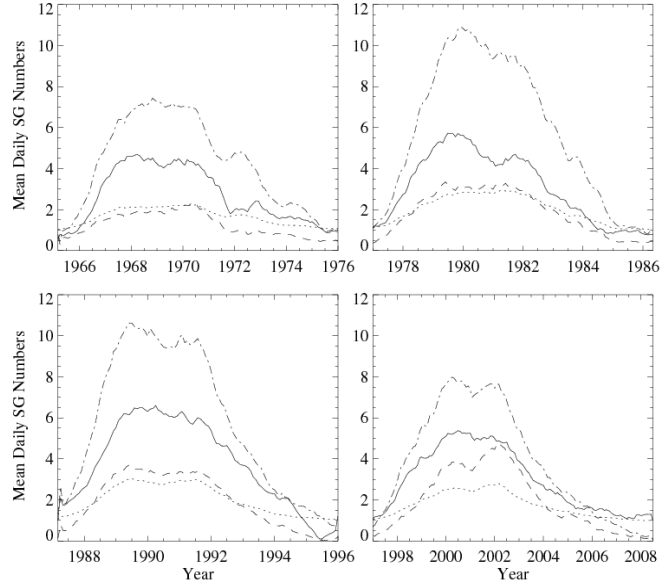

*Figure 3. Solar cycle variations of selected parameters smoothed with a 12 step running average filter. For display purposes, the ISSN and 10.7 cm solar radio data were re-scaled: the ISSN was divided by 15, while the F10.7 cm radio flux is divided by 700. The solid (dashed) line represents small (large) SG numbers, dashed–dotted line is ISSN, and dotted line is F10.7 radio flux.*

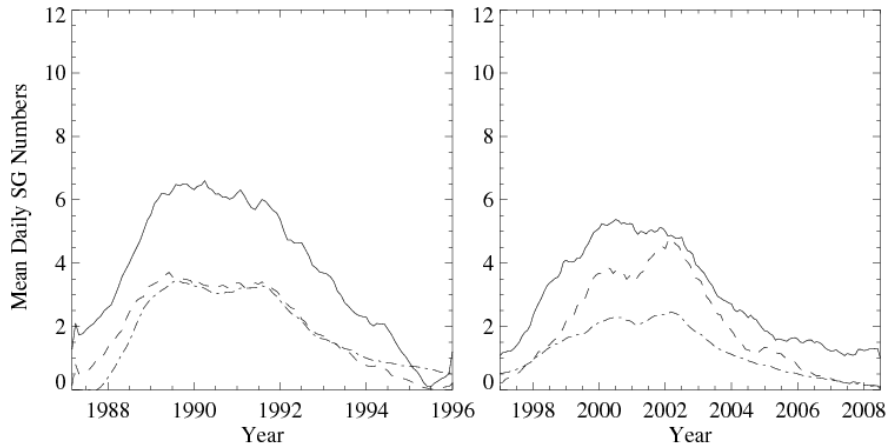

*Figure 4. Facular area (dash-dotted line) compared with time variations of small (solid) and large (dashed) SG numbers during solar cycles 22 (left) and 23 (right). All data points were smoothed with a 12 step running average filter. For the display purposes the facular area was divided by 20000.*

Figure 5 shows the SG numbers together with measurements of the maximum CME speed index introduced by Kilcik et al. (2011). This index is a maximum CME speed index computed by averaging daily maximum CME speeds over a one month period. The CME speed data are the linear fit speeds from the CME catalog[2] (Yashiro et al. 2004; Gopalswamy et al. 2009), and they are available for 23$^{rd}$ solar cycle only.

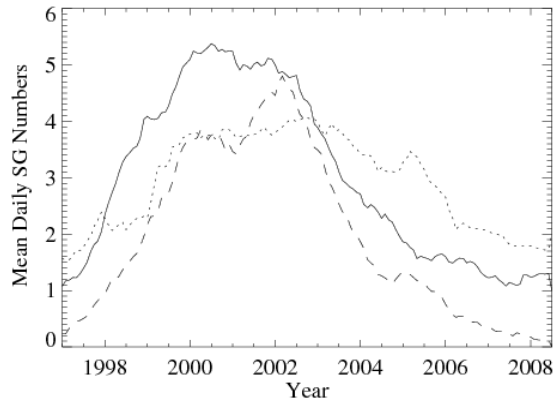

*Figure 5. Solar cycle variations of the monthly maximum CME speed (dotted line), large (dashed line) and small (solid) SG numbers. All data were smoothed with a 12 step running average filter. For display purposes CME speed data are re-scaled by dividing by 200.*

There are two points we would like to address in Figure 5. First, in the beginning of the declining phase of the small SG numbers (after year 2000), the maximum CME speed index still continued to increase and it reached its highest point approximately 6 months after the peak in the large SG numbers. Second, there is a local peak in the maximum CME speed index time profile centered at approximately April 2005, and this peak is co-temporal with the enhanced large SG numbers (dashed line). Thus, we conclude that the maximum of the CME speed index shows a better agreement with the large SG numbers. Also the cross-correlation coefficient between the CME speed index and the large SG numbers is greater as compared to the small SG numbers (see Table 1 bottom row).

*Table 1. Cross–correlation and Fisher's test analysis of large and small AR counts with other data sets used in this study* Table 1

| Cycle # / Data | Cycle 20 | | Cycle 21 | | Cycle 22 | | Cycle 23 | |
|---|---|---|---|---|---|---|---|---|
| | Small ARs | Large ARs | Small ARs | Large ARs | Small ARs | Large ARs | Small ARs | Large ARs |
| ISSN | 0.85 ± 0.04/0.05 | 0.82 ± 0.05/0.06 | 0.90 ± 0.03/0.04 | 0.92 ± 0.02/0.03 | 0.92 ± 0.02/0.03 | 0.97 ± 0.01/0.02 | 0.93 ± 0.02/0.02 | 0.95 ± 0.01/0.02 |
| F10.7 | 0.84 ± 0.04/0.06 | 0.80 ± 0.05/0.07 | 0.87 ± 0.04/0.05 | 0.91 ± 0.03/0.04 | 0.90 ± 0.03/0.04 | 0.96 ± 0.01/0.02 | 0.88 ± 0.03/0.04 | 0.95 ± 0.01/0.02 |
| Facular Area | ⎯⎯⎯ | ⎯⎯⎯ | ⎯⎯⎯ | ⎯⎯⎯ | 0.88 ± 0.04/0.05 | 0.89 ± 0.03/0.04 | 0.91 ± 0.02/0.03 | 0.94 ± 0.02/0.02 |
| CME index | ⎯⎯⎯ | ⎯⎯⎯ | ⎯⎯⎯ | ⎯⎯⎯ | ⎯⎯⎯ | ⎯⎯⎯ | 0.64 ± 0.09/0.10 | 0.70 ± 0.07/0.09 |

---

[2] http://cdaw.gsfc.nasa.gov/CME_list/index.html

We use the information presented in Figures 2-5 to quantify the relationship between the SG numbers and other solar activity indicators (ISSN, F10.7 cm solar radio flux, facular area and CME speed index). We note that the cross-correlation analysis was applied to unfiltered daily mean values averaged over one month. Fisher's test was further utilized to estimate the error level in the cross-correlation coefficients and to obtain upper and lower bounds in the 95 % in confidence. We find that, compared to the small SG number, the large SG number better correlates with the 10.7 cm solar radio flux (except in cycle 20), the facular area, and the maximum CME speed index (see Figure 6). Both the small and large SG numbers correlate equally well with the ISSN. Although the difference between the cross-correlation coefficients of large/small SG numbers and other indicators is not generally significant, the large SG numbers generally display a higher cross-correlation with other parameters, when compared to small SG cross-correlation coefficients.

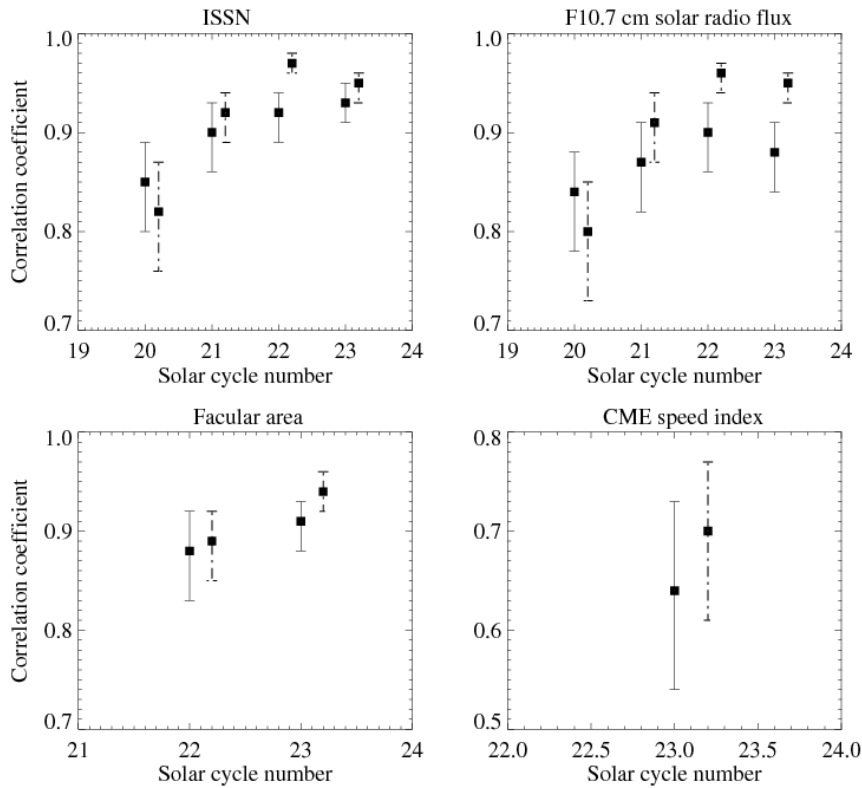

*Figure 6. Fisher's test plots. Each panel shows the cross–correlation coefficients and Fisher's test results of comparison of large (dashed) and small (solid) SG numbers with other data sets. The black squares show the cross-correlation coefficients*

In Figure 7, we plot the monthly total sunspot area (light gray) and the monthly total sunspot number (dark gray), as well as their ratio (or average area of a sunspot, thick line) as inferred from LEAR data. The plotted data are only for solar cycles 22 and 23, and they are presented separately for large (upper panel) and small (lower panel) SGs. As is evident from the lower panel, the total area of small sunspots in cycle 22 by far exceeds that for solar cycle 23, this trend is also seen in the SG number plots (Figure 3). The averaged area of small sunspots (lower panel, thick line) was nearly the same in both solar cycles. In contrast, while the total area of large sunspots was larger during cycle 22 as compared to cycle 23, the total number of large SGs (and the daily large SG numbers) and are nearly the same. As a result, the average large sunspot area is about 25% smaller in solar cycle 23 as compared to cycle 22. In other words, there were slightly more D, E, and F sunspots in cycle 23 but they were, on average, than for to solar cycle 22.

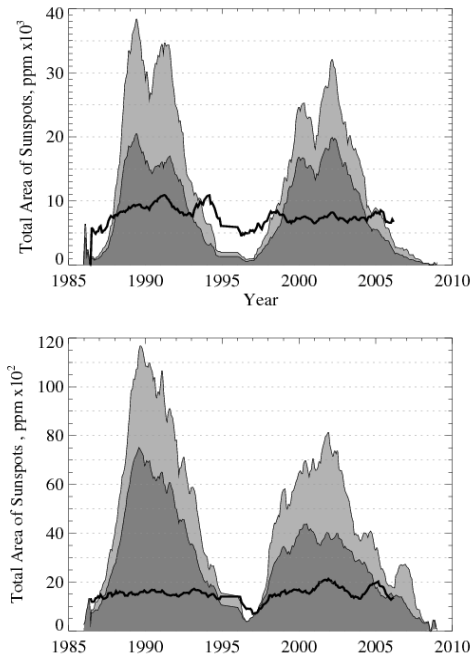

*Figure 7. Time variation of monthly total sunspot areas (light gray), monthly total number of sunspots (dark gray) and the average area of a sunspot (thick line). The plotted data are only for solar cycles 22 and 23, and it is presented separately for large (upper panel) and small (lower panel) SGs.*

To better represent the time distribution of the large groups through the investigated solar cycles, we created a phase plot (Figure 8), in which we plot the ratio of the large SG numbers to the total SG numbers (small plus large) versus solar cycle phase. To calculate the phase of the solar cycle, we first take the length of a solar cycle to be unity and then express each point in time as a fraction of the cycle (unity) for each cycle, separately. It is well known that the ascending phase of the sunspot cycle is much shorter than the descending phase[3.] The corresponding mean values determined for solar cycles 20-23 are 3.6 and 7.6 years[3], respectively. The mean phase of the ascending time was found to be about 0.32. The ascending phase was calculated as the ratio of the mean ascending time to the mean length of the cycle in question. As follows from Figure 8, the large groups become most dominant at phase of 0.46 (nearly the middle of a cycle), which is much later than the ISSN maximum (phase 0.29-0.35, depending on the solar cycle). The shortest (in our data set) cycle is 22, which displays a distinct behavior from other cycles. While it does have a maximum at phase 0.46 as the other cycles do, this maximum is not well–pronounced and in general, we may conclude that the time profile of the large SG numbers for cycle 22 is nearly the same as for the small SG numbers. On the other hand, the most unusual is the solar cycle 23, which not only has the highest large SG numbers, but also a very prominent peak at phase 0.46. It also has two more local peaks at phases 0.3 (year 2000) and 0.71 (year 2005). These two local maxima are contemporaneous with the maximum of the ISSN and the enhanced CME occurrence in 2005.

---

[3] ftp://ftp.ngdc.noaa.gov/STP/SOLAR_DATA/SUNSPOT_NUMBERS/docs/maxmin.new

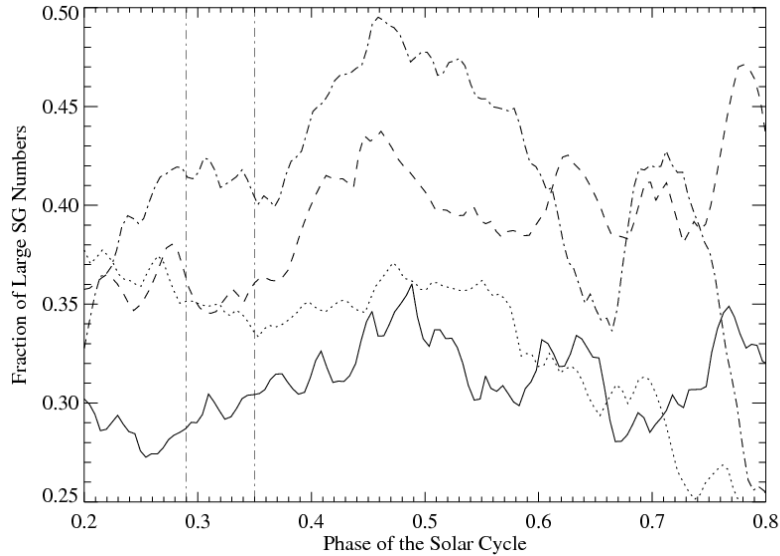

*Figure 8. The ratio of the large SG number to the total number of all active regions (i.e., sum of small and large SG numbers) plotted versus the solar cycle phase. Solid line is cycle 20, dashed line is cycle 21, dotted line is cycle 22, and dashed dotted line is cycle 23. Vertical lines indicate the range of the maximum phase of the ISSN determined for the four studied solar cycles.*

3. Conclusions and Discussion

In this study, we analyzed averaged daily SG numbers separately calculated for large and small SGs as observed during solar cycles 20-23. The main findings of this study are as follows:
1) The time distributions of the large and small SG numbers show asymmetries except for cycle 22. While the number of small SG peaks at the time of the first ISSN maximum, the peak of the large groups is delayed by about of 2 years and is co-temporal with the second ISSN maximum.
2) Large SGs tend to reach their maxima nearly half-way through the solar cycle (phase 0.45-0.5), while the ISSN generally peaks at solar cycle phase of 0.29-0.35.
3) During the most recent cycle, 23, the large SG numbers were higher (and the total number of sunspots belonging to large SGs was nearly the same) when compared to those for solar cycle 22. At the same time the total sunspot area in cycle 23 was smaller.
4) The F10.7 radio flux, facular areas and the maximum CME speed show better agreement with the large SG numbers than they do with the small SG numbers.

We begin the discussion by pointing out that Gopalswamy et al. (2003) reported that for solar cycle 23, the CMEs occurrence rate peaked two years after the solar maximum, as determined from the ISSN. Later, Kilcik et al. (2011) found that the monthly averaged maximal speeds of CMEs lag behind the sunspot numbers by several months. These two studies are in line with our findings that the peak of the large SGs count is delayed relative to the solar cycle maximum. Further, Gopalswamy et al. (2003) separated all CME events into two classes according to the location of their source region and reported that high-latitude CMEs ceased by the first quarter of 2002, which implies that the delayed excess of CMEs during the declining phase of the solar cycle (2003-2005) was caused by low-latitude CMEs related to both filaments and flares. While Joshi & Pant (2004) found that Hα flare activity between 1996 and 2003 was low compared to previous solar cycles, Bai (2006), in turn, analyzed distribution of X-ray solar flares and reported that the weaker solar cycles 20 and 23 produced larger number of major flares in their declining phases, as compared to stronger solar cycles 21 and 22. Bai's (2006) report agrees well with our findings on the domination of large groups in the declining phase and may suggest, according to Bai, that the Sun could became more efficient in transporting magnetic fields from the base of convection zone to the surface during the declining phase of solar cycle.

As we stated earlier in the text, the duration of the ascending phase of the solar cycle is remarkably shorter than that of the descending phase. Our data suggest that this rule is not obeyed in the case of large SGs: they peak at phase of about 0.5, regardless of the cycle length, while ISSN and the small groups (see Figure 3) reach their maximum much earlier in the phase of the cycle at about 0.32. Thus, small groups seem to have a dominant effect on the ISSN in the early phase of the solar cycle. We thus arrive at the conclusion that, at least during the descending phase of a solar cycle, the number of large SGs seems to be a better descriptor of CME and flare activity (see Figure 8 dashed – dotted line). Furthermore, since fast CMEs largely define geomagnetic activity (Kilcik et al. 2011), the large SG numbers may also be more valuable in describing solar-terrestrial relationship, as compared to the ISSN.

De Toma et al. (2004) compared the TSI, magnetic flux, SSA, facular area, composite Mg II 280 nm index and F10.7 flux. The authors showed that, similar to our conclusions, all these parameters reached their maximum in the year 2002 (two years later than the ISSN). Also, similar to the ISSN, they all show a double peak structure near the maximum of the solar cycle (see Figure 1 in de Toma et al. 2004). We argue that this two year time delay may be, at least partially, due to an excess of large SGs present on the solar disk during the second half of cycle 23.

The TSI values measured near the maximum phase of solar cycle 23 are comparable to these from solar cycle 22, in spite a significant difference in their ISSN magnitudes. De Toma et al. (2004) thus concluded that the higher than expected magnitude of the TSI may be caused by a combined action of lower sunspot area and an excess of relatively weak SGs, and therefore, larger plage regions. However, Figure 4 shows that the facular area is lower in cycle 23 than in cycle 22 and it better correlated with the large SG numbers as compared to the small SG numbers. The TSI profile for cycle 23 has a double peak and the second peak begins in the middle of 2001 (see Fig. 1 in de Toma et al. 2004) and it approximately ends in the middle of 2002, thus coinciding with the peak of the large SG numbers. Next, the small SG numbers during solar cycle 22 are significantly *higher* than there in solar cycle 23, while the large SG numbers show the opposite ratio (also see Figure 3). The decrease of the area ratio of faculae and sunspots with increasing activity has been reported earlier (see Foukal 1998 for details). It appears that the decrease of both facular area and the number of small SGs in the weaker cycle 23 may have been compensated by the higher number of large SGs. They may have brighter faculae, so that the resulting TSI levels remained high. Ortiz et al. (2002) analysis supports this suggestion. These authors showed that in general, faculae contrast grows with increasing field strength. The weak field (network) faculae have low contrast that only weakly depends on the heliocentric angle. On the other hand, strong field faculae display noticeably higher contrast at larger heliocentric angles (closer to the solar limb), while the contrast decreases at the disk center and even becomes negative for very strong fields (i.e., micropores, see also Spruit 1976, Lawrence et al. 1993 and Topka et al. 1997). The dependence of contrast on the magnetic field strength (and the size of a flux tube) may indicate that the size distribution of flux tubes in network and active region faculae may be different (see Keller 1992, Grossmann-Doerth et al. 1994).

Also, as we discussed above, the average area of a sunspot belonging to a large SG decreased in solar cycle 23. Our interpretation of this somewhat unexpected finding is that the large SGs of cycle 23 could be more complex, i.e., they may have displayed a greater number of small sunspots/pores, which added significantly to the total number of sunspots, but their contribution to the total sunspot area, because of their size, was not significant.

The above discussion was based on reliable data for the only two solar cycles. The SG numbers for cycles 20 and 21 are based on data from the ROME station and suffer from many data gaps causing the ROME counts to be systematically lower than the LEAR data. However, as we discussed earlier, ROME large (small) SG numbers need to be multiplied by 1.2 (1.6) to match the LEAR counts. When we correct the ROME data by using these scaling coefficients we find that large and small SG numbers for cycle 21 are higher than those for cycle 22. The maximum TSI level for cycle 21 is also higher as compared to cycle 22 (Fröhlich 2009).

One indirect confirmation of the role that large SG may play in contributing to the TSI level comes from the work by Wenzler et al. (2005), who reconstructed the TSI by using spectromagnetograph data obtained at the National Solar Observatory, Kitt Peak between 1992 and 2003. These authors found nearly same level of the TSI for cycles 22 and 23. They also argued that the TSI variations were mainly caused by the evolution of the solar surface magnetic fields, which may in turn be linked to sunspot groups, whose number (in the second half of the cycle) was largely determined by the occurrence of large groups. Frohlich (2009) reached the opposite conclusion and reported that the long-term change in TSI is caused by a global temperature change of about 0.2 K during cycle 23. The authors further argue that the UV and EUV irradiances do not show any long-term trend. Our analysis seems to support the conclusions of Wenzler et al. (2005). Advancing our understanding of the nature of long term variations of TSI is not only important for solar variability studies, but is also vital for the solar-terrestrial connections.

We would like to thank the referee for valuable comments and suggestions, which led to a significant improvement of the paper. We acknowledge usage of ISSN, sunspot groups, solar cycle length and 10.7 cm solar radio flux from National Geophysical Data Center. We acknowledge that the facular area data from the San Fernando Observatory, California State University Northridge and the F10.7 radio flux data was obtained at the Dominion Radio Astrophysical Observatory, National Research Council of Canada. This research was supported by NASA grants GI NNX08AJ20G and LWS NNX08AQ89G as well as NSF ATM0716512 grant.